\documentclass[pmlr,twocolumn,10pt]{jmlr} % W&CP article

\usepackage{booktabs}
\usepackage{siunitx}
\theorembodyfont{\upshape}
\theoremheaderfont{\scshape}
\theorempostheader{:}
\theoremsep{\newline}

\title{Few-Shot Transfer Learning to improve Chest X-Ray pathology Detection using limited triplets}

\author{%
\Name{Ananth Reddy Bhimireddy} \Email{ananth.reddy@emory.edu}\\
\addr Emory University, USA
\AND
\Name{John Lee Burns} \Email{jolburns@iupui.edu}\\
\addr Indiana University Purdue University-Indianapolis, USA
\AND
\Name{Saptarshi Purkayastha} \Email{saptpurk@iupui.edu}\\
\addr Indiana University Purdue University-Indianapolis, USA
\AND
\Name{Judy Wawira Gichoya} \Email{ judywawira@emory.edu}\\
\addr Emory University, USA
\AND
}

\begin{document}

\maketitle

\begin{abstract}
Deep learning approaches applied to medical imaging have reached near-human or better-than-human performance on many diagnostic tasks. For instance, the CheXpert competition on detecting pathologies in chest x-rays has shown excellent multi-class classification performance. However, training and validating deep learning models require extensive collections of images and still produce false inferences, as identified by a human-in-the-loop. In this paper, we introduce a practical approach to improve the predictions of a pre-trained model through Few-Shot Learning (FSL). After training and validating a model, a small number of false inference images are collected to retrain the model using \textbf{\textit{Image Triplets}} - a false positive or false negative, a true positive, and a true negative. The retrained FSL model produces considerable gains in performance with only a few epochs and few images. In addition, FSL opens rapid retraining opportunities for human-in-the-loop systems, where a radiologist can relabel false inferences, and the model can be quickly retrained. We compare our retrained model performance with existing FSL approaches in medical imaging that train and evaluate models at once.

\end{abstract}

\paragraph*{Data and Code Availability} \label{data_code_availability}

This paper uses CheXpert Dataset \cite{DBLP:journals/corr/abs-1901-07031}, which is a publicly available dataset of 224,316 chest radiographs from 65,240 patients.

The code used for experimentation processes described in this paper - which include triplet dataset creation, training and validation of the Few-Shot Learning and Incremental Few-Shot Learning Models were anonymized and included in the supplemental material section. 

% This initial paragraph is \textbf{mandatory}. Briefly state what data you
% use (including citations if appropriate) and whether the data are
% available to other researchers.\footnote{An example data availability
% statement: This paper uses the MIMIC-III dataset
% \citep{johnson2016mimic}, which is available on the PhysioNet repository
% \citep{johnson2016physionet}.}
% If you are not sharing code, you must explicitly state that you are not
% making your code available. If you are making your code available, then
% at the time of submission for review, please include your code as
% supplemental material or as a code repository link; in either case, your
% code must be anonymized. If your paper is accepted, then you should
% de-anonymize your code for the camera-ready version of the paper. \emph{If
% you do not include this data and code availability statement for your
% paper, or you provide code that is not anonymized at the time of
% submission, then your paper will be desk-rejected.} Your experiments later
% could refer to this initial data and code availability statement if it is
% helpful (e.g., to avoid restating what data you use).

\section{Introduction}
\label{sec:intro}

Improvements in deep learning algorithms and availability of large annotated datasets have been critical in the gains achieved in computer vision tasks including segmentation and classification. However, the application of these algorithms remains a challenge for medical imaging, as computer vision algorithms usually require a large amount of well-annotated datasets \cite{rajpurkar2017chexnet}. In medical practice, annotated datasets are expensive to curate \cite{doi:10.1148/radiol.2020192224}, limited by HIPAA/GDPR and other regulations \cite{10.1001/archinte.165.10.1125}, and often focused on a specific medical condition limiting generalizability. 

Deep learning approaches that perform classification and segmentation tasks require large annotated datasets. Usually, a deep learning neural network will go over this large dataset multiple times (called epochs) to continuously adjust the weights of the network nodes during the training process. A contrarian approach to this traditional (big-shot) approach, which many researchers are working on under the terms, few-shot learning (FSL), one-shot learning, zero-shot learning \cite{peng2015learning}, is to use fewer epochs or fewer images in the neural network training \cite{ravi2016optimization}. In medical imaging, the big-shot approach is tedious, labor-intensive, and thus expensive because radiologists' time is costly \cite{chartrand2017deep}.

\begin{figure*}[!bp]
   \centering
   \includegraphics[width=0.75 \linewidth]{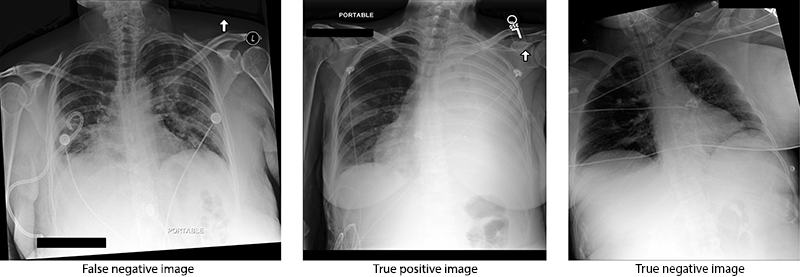}
   \caption{Pneumothorax detection - triplets}
   \label{Fig:Triplets}
\end{figure*}

Few-Shot Learning on label limited datasets in medical imaging is not new \cite{pmlr-v106-prabhu19a, medela2019few, paul2021discriminative}. In our proposed approach, we work with limited data and inter-annotator differences. We evaluate our approach using the CheXpert \cite{DBLP:journals/corr/abs-1901-07031} Chest X-rays dataset. Our approach of blending an FSL algorithm with image triplets, which we call Triplet Few-Shot Learning (TFSL), is practical and novel. \textbf{\textit{Image Triplets}} is a set of three images - a false positive (FP) or false negative (FN) generated by the model, a true positive (TP), and a true negative (TN). 

% We implemented the algorithm with a limited number of image triplets that were curated to avoid data leakage. Algorithm validation is performed on the false inference images using the image triplets. The model is retrained using the Adam optimizer and Margin Ranking Loss function. Usual classification metrics such F1-Score, precision and recall are not reported, as these metrics do not convey changes in false positive and false negative. We evaluated performance using Positive Prediction Value (PPV) and Negative Prediction Value (NPV) metrics.  All image triplets, including all pathologies within CheXpert, are trained and validated at the same time. Experiments were ran to evaluate the improvement in training time and metrics of the Triplet Few-Shot Learning (TFSL) and Incremental FSL models. TFSL re-training is rapid, and can use human-in-the-loop annotations to retrain with new labels. No other existing model can aid in these kinds of applications.

\section{Related Works}

Various deep learning techniques including Convolutional Neural Networks (CNN), Long-Shot Term Memory Networks, fast AnoGAN (f-AnoGAN), and Multi-model fusion strategies are  applied to medical imaging tasks \cite{gao2018fully,du2016overview,DBLP:journals/corr/KamnitsasLNSKMR16,DBLP:journals/corr/abs-1808-01200,SCHLEGL201930}. Regardless of the deep learning technique applied, large annotated data sets are required to train AI algorithms, and the model outputs often include false inferences.FSL is applied in computer vision detection and segmentation tasks allowing robust model generation from small data sets, as extreme as One Shot Learning utilizing one positive/negative image pair \cite{chanda2019face, chen2019selfsupervised, hu2020leveraging, luo2019fewshot}

FSL is a form of meta-learning which allows generalization to new classes on small data sets. FSL has been applied in medical imaging for many tasks. In an ISBI 2019 paper, Medela et al. applied FSL to reduce the need for labeled biological datasets in histopathology \cite{medela2019few}. Ali et al. (2020) applied it for classifying endoscopy images \cite{ali2020additive}. More recently, FSL has been applied to interactive training during segmentation, similar to our vision of using it in human-in-the-loop situations \cite{feng2021interactive}, in COVID-19 diagnosis from CT scans \cite{chen2021momentum}, and detecting rare pathologies from fundus images that were collected for a different purpose like Diabetic Retinopathy \cite{quellec2020automatic}. FSL can perform better than other frameworks such as transfer learning and Siamese networks to detect rare conditions usually represented with few images \cite{quellec2020automatic, feyjie2020semi, kim2017few}.

\section{Technical Description}

The TFSL algorithm is designed using MarginRankingLoss to reduce the number of false inferences made by the model. The TFSL algorithm is built on the best-performing pre-trained model submitted to the CheXpert competition. The pre-trained model used in the paper for experiments is trained using CheXpert data and is available at - \url{https://github.com/gaetandi/cheXpert}. An inference of this model is run to create a baseline model.

\subsection{Triplet Data Creation}

CheXpert dataset is used to train and evaluate the TFSL approach on image triplets. The first image in the image triplet is randomly selected from the failed inference images of the baseline model. Theoretically, the inference failure can be either an FP or FN. The second image in the image triplet is a TP and the third image is a TN.

While collecting the \textit{training image triplets} (Fig \ref{Fig:Triplets}), a checking label is also collected. The checking label is -1 if the first image of the triplet is FN and 1 if the random image is an FP. The choice of -1 and 1 as labels is explained in Section \ref{modeling-choices}. We tested sets of 50/100/150 image triplets and the fine-tuning model improved performance over the baseline model at 150 image triplets. The randomly selected 150 image triplets were used for training the TFSL algorithm, but all the failed inference images (except the ones that are used for training) were collected and used to validate the algorithm. False inference images are randomly selected from all the failed inference images.

\subsection{Baseline Model}

The pre-trained model uses a DenseNet121 algorithm to classify pathologies from an image. Before training, images are converted to RGB, resized to (320x320) and further cropped to 224. The image data is converted into a PyTorch DataLoader. Adam optimizer and BCELoss (Binary Cross-Entropy Loss) are used to build the pre-trained model. Inference (evaluation) of the pre-trained model was used as the baseline model in this paper. We also implemented the FSL algorithm by following the guidelines outlined \cite{cermelli2020guidelines} and refer to it as \textbf{Incremental FSL Training} in this paper.

\subsection{Triplet Few-Shot Learning (TFSL) Model}

The pre-trained classification model was modified by replacing the final layer (Linear layer and Sigmoid activation function) with a Linear layer of 128 units and PReLU (Parametric Rectified Linear Unit) \cite{he2015delving} activation function to create 128-dimensional vectors for every image in the image triplets. The architecture of data set creation and modeling the TFSL algorithm is summarized in Fig \ref{Fig:Architecture}.

\begin{minipage}[b]{0.9\linewidth}
    \centering
    \centerline{\includegraphics[width=8.5cm]{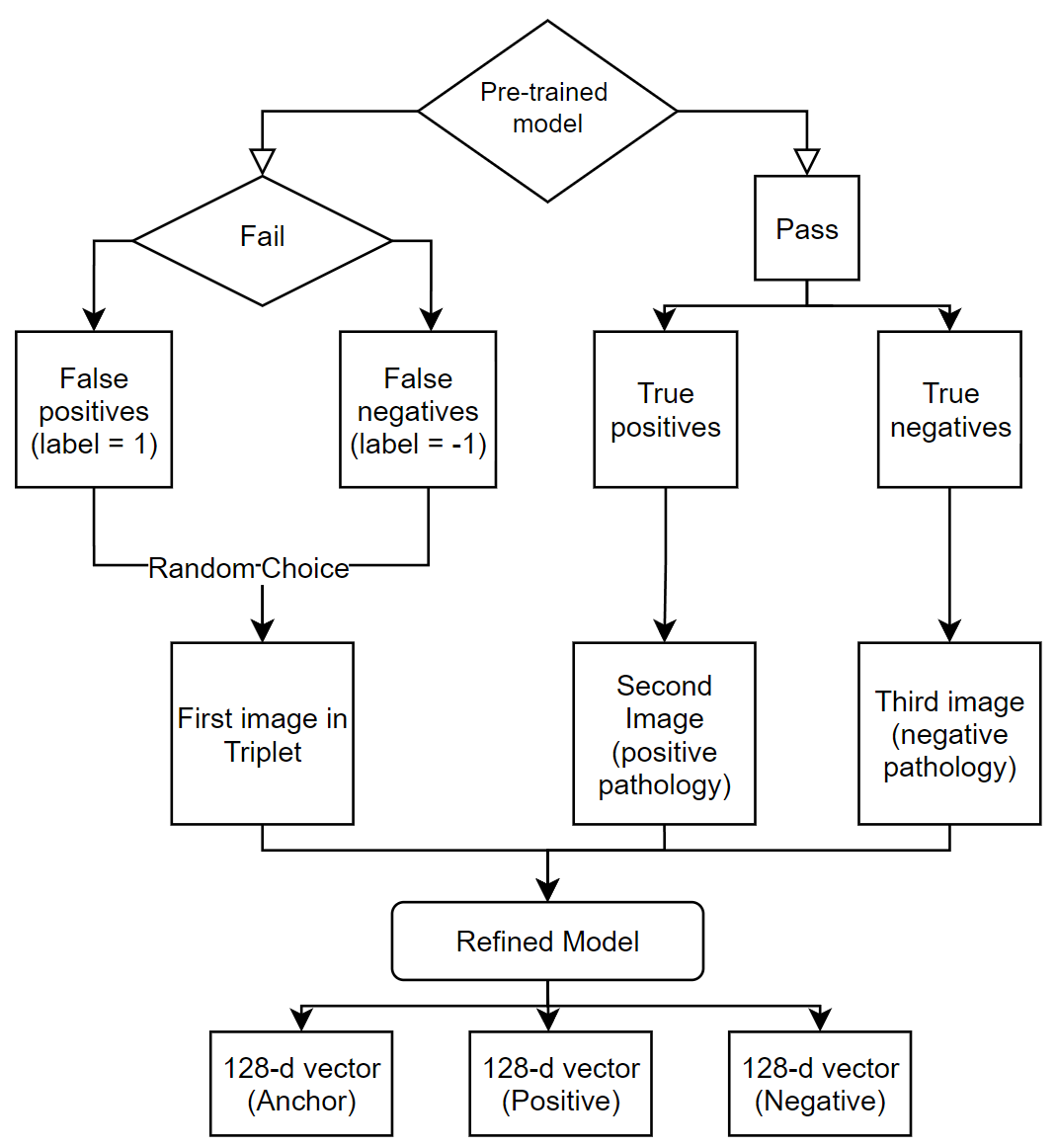}}
    \label{Fig:Architecture}
    % \caption{Process Architecture}
\end{minipage}

Each image in the image triplet was transformed into a 128-dimensional vector. These 128-dimensional vectors are used to calculate the distance between the images. If the false inference image is FN, then the image should be closer to the TP image from the triplet in an n-dimensional space; conversely if the false inference is FP, then the image should be closer to the TN image from the triplet in an n-dimensional space. We use the pre-trained classification model to create the n-dimensional vectors. 

The TFSL model is trained for five epochs using Adam Optimizer, a learning rate of 0.0001, and a weight decay of 1e-5. Positive Predictive Value (PPV) and Negative Predictive Value (NPV) are used as evaluation metrics. Margin Ranking Loss is used as the loss function. The training approach is slightly different from the Incremental FSL approach, in which the model was trained and validated on all 14 pathologies at once.

\subsection{Modeling Choices} \label{modeling-choices}

Margin Ranking Loss function takes two inputs x1, x2 and returns 1 if the first input is ranked higher and returns -1 if the first input is ranked lower. We chose -1 and 1 as labels while creating training image triplets to mimic this pattern. The euclidean distance between first image vector and second image vector will be x1 and the euclidean distance between first image vector and third image vector will be x2. The mathematical form of Margin Ranking Loss function (\ref{MRLF_Formula}) is provided below - where x1, x2 are the loss inputs and y is the label tensor. 

\begin{equation}
    \label{MRLF_Formula}
    loss(x1, x2, y) = max(0, -y*(x1-x2) + margin)
\end{equation}

PPV and NPV are chosen as evaluation metrics for the FSL because they aid in identifying the decrease in False Positives and False Negatives, respectively. PPV increases as the FPs are converted to TNs. NPV increases as FNs are converted to TPs.

\section{Experiments and Results} \label{sec:experiments}

 All the models designed, trained, and evaluated in this paper uses the PyTorch. We open-sourced the dataset creation, training and validation code \footnote{The code for this paper is at - \url{https://github.com/iupui-soic/Radiology_FSL}}. We compare the baseline DenseNet Model to TFSL model and incremental FSL models. The results are provided in Table \ref{Results}.

\begin{table}[!ht]
\small
\vspace{-2mm}
\caption{Results from the Baseline, TFSL and Incremental FSL Models}\label{Results}
\setlength{\tabcolsep}{0.5pt}
\begin{center}
\vspace{-4mm}
\begin{tabular}{|*{8}{c|}}
\hline
\multicolumn{1}{|c}{}& \multicolumn{2}{|c|}{\shortstack{\\\textbf{Baseline} \\ \textbf{Model}}} & \multicolumn{2}{|c|}{\shortstack{\\\textbf{TFSL} \\ \textbf{Model} }} & \multicolumn{2}{|c|}{\shortstack{\\\textbf{Incremental} \\ \textbf{FSL Model}}}\\ \hline

\multicolumn{1}{|c}{Pathology} & \multicolumn{1}{|c}{PPV} & \multicolumn{1}{|c}{NPV} & \multicolumn{1}{|c}{PPV} & \multicolumn{1}{|c}{NPV} & \multicolumn{1}{|c}{PPV} & \multicolumn{1}{|c|}{NPV} \\ \hline

No Finding & 89.99 & 10.35 & 94.06 & 44.38 & \textbf{94.53} & \textbf{47.91} \\ \hline
\shortstack{\\Enlarged \\ Cardiomediastinum} & 100.0 & 0.0 & 100.0 & 40.0 & 100.0 & \textbf{53.28} \\ \hline
Cardiomegaly & 88.63 & 11.30 & 93.88 & 47.87 & \textbf{93.93} & \textbf{51.32} \\ \hline
Lung Opacity & 71.83 & 28.51 & \textbf{86.14} & 61.23 & 85.53 & \textbf{63.56} \\ \hline
Lung Lesion & 99.91 & 0.0 & 99.97 & 44.14 & 99.98 & \textbf{48.16} \\ \hline
Edema & 80.49 & 20.01 & 89.33 & 50.49 & \textbf{89.80} & \textbf{54.14} \\ \hline
Consolidation & 99.98 & 0.0 & 100.0 & 42.79 & 100.0 & \textbf{46.97} \\ \hline
Pneumonia & 99.48 & 0.57 & \textbf{99.82} & \textbf{48.48} & 99.70 & 47.69 \\ \hline
Atelectasis & 94.48 & 5.74 & \textbf{97.31} & 47.29 & 97.11 & \textbf{50.04} \\ \hline
Pneumothorax & 99.92 & 0.0 & \textbf{99.94} & 42.46 & 99.90 & \textbf{48.64} \\ \hline
Pleural Effusion & 79.94 & 20.25 & \textbf{89.70} & 52.08 & 88.56 & \textbf{55.85}  \\ \hline
Pleural Other & 100.0 & 0.0 & 100.0 & \textbf{49.47} & 100.0 & 46.31 \\ \hline
Fracture & 100.0 & 0.0 & 100.0 & 50.0 & 100.0 & \textbf{50.73} \\ \hline
Support Devices & 75.19 & 23.31 & \textbf{87.49} & 57.63 & 86.96 & \textbf{58.13} \\ \hline
\end{tabular}
\end{center}
\vspace{-2mm}
\end{table}

\section{Results} \label{sec:results}

The baseline model has high PPV and low NPV values. The TFSL model reduced FP and FN outputs, indicated by an increase in PPV and NPV, respectively. By performing a statistical test, we concluded that the TFSL algorithm and Incremental FSL algorithm results improved over the baseline results. After performing a similar test on TFSL and Incremental FSL algorithms, we concluded that the PPV values did not significantly differ from each other. The Incremental FSL NPV was higher than both the TFSL model with a statistical significance value (p-value) of 0.007. We used the dependent t-test model from \textit{statsmodel} in the \textit{Scipy} package to perform the above statistical tests.

The time taken to train and validate the TFSL model on any pathology is around 8-9 minutes on a Nvidia Quadro P6000 GPU with 24 GB memory. This provides us with the ability to label and train the model rapidly. The Incremental FSL also consumed the same amount (8-9 minutes) to train and validate the algorithm on any pathology. 

\section{Discussion} \label{sec:discussion}

FSL architectures are growing within medical imaging \cite{PAUL2021101911}. This work builds on FSL architecture through the use of triplets as shown in Figure 2.  TFSL requires less training data and time compared to previous approaches. While a two-step process of using a saliency-based classifier with a discriminative autoencoder ensemble \cite{PAUL2021101911} has better performance in FSL compared to our approach, the simplicity and speed of our approach are important advantages to be considered. Our approach for fine-tuning models can be taken to edge devices, as has been shown in the non-medical imaging domain \cite{lungu2020siamese}.

Additionally, all previous FSL architectures consider a singular ground truth label for images. Image labeling can have variability, particularly among radiologists when annotating studies \cite{cabitza_bridging_2020,saha_breast_2018}. Our approach is able to use triplets that are determined by human-in-the-loop's annotations and train a model that is specific to these new labels. Thus, the TFSL approach can deal with rapid re-training required on false inference images as determined by the user radiologist. The baseline model failed to identify true negative inference images frequently in pathologies such as Enlarged Cardiomediastinum, Lung Lesion, Consolidation, Pleural Other and Fracture, and at other times failed to identify true positives. Fine-tuning in TFSL and Incremental FSL substantially improved the identification of TP and TN inference images. A TFSL and human-in-the-loop system can be retrained easily through transfer learning methods made difficult by other approaches.

\section{Conclusion} \label{sec:conclusion}

In the paper, we presented a comparison of results between baseline and fine-tuned models, providing a conclusive evidence that the TFSL algorithm outperformed the baseline model. The use of Margin Ranking as the loss function, performance gains in limited datasets with quick retraining, and the simplicity of our approach are important characteristics of the TFSL approach. In the future, we plan to test this approach on different modalities and non-medical/natural image data sets.

In summary, the major contributions of our paper are as follows:
\begin{enumerate}
\item We present a modified Few-Shot Learning algorithm to effectively improve the results of predicting pathologies on images whose inferences failed.
\item We present a comparison of results between a fine-tuned model,  our Few-Shot Learning model and a previously published few short learning algorithm trained in an incremental fashion. Our model out-performs the fine-tuned model and achieves a higher NPV in all classes, with close performance to the incremental FSL. We demonstrate that we can get good performance at lower computation.  
 \item We experiment with MarginRankingLoss and TripletMarginLoss function as loss functions. Despite the assumption that TripletMarginLoss would perform better for image triplets, we found that MarginRankingLoss is more appropriate for our use case. This is not described in any of the previous works.
 \item Previous works have evaluated Few-Shot Learning on fewer classes within one or multiple datasets. We present our experiments using the CheXpert dataset to improve the predictions on 14 pathologies on Chest X-rays.
\end{enumerate}

\section*{Institutional Review Board (IRB)}

This research work performed in this paper does not require IRB approval as the data is an open-source dataset.

% This section is \textbf{mandatory}. If your research requires IRB
% approval or has been designated by your IRB as Not Human Subject
% Research, then for the camera-ready version of the paper, you must
% provide IRB information (and at the time of submission for review, you
% can say that this IRB information will be provided if the paper is
% accepted). If your research does not require IRB approval, then you
% must state this to be the case. This section does not count toward the
% paper page limit.

% \acks{Acknowledgments go here \emph{but should only appear in the
% camera-ready version of the paper if it is accepted}.
% Acknowledgments do not count toward the paper page limit.}

\bibliography{jmlr-sample}

\end{document}